\documentclass{iopart}
\usepackage{graphicx}  
\usepackage{latexsym}  

\jl{6}        % submitted to journal: Classical and Quantum Gravity
\eqnobysec    % section equation numbering

\def\beq{\begin{equation}}
\def\eeq{\end{equation}}

\def\rmd{{\rm d}}

\begin{document}

\title[Motion of extended bodies in Schwarzschild spacetime]
{Quadrupole effects on the motion of extended bodies in Schwarzschild spacetime}

\author{
Donato Bini$^* {}^\S{}^\P$, Pierluigi Fortini$^\dagger$,
Andrea Geralico$^\ddag {}^\S$ and Antonello Ortolan$^{**}$}
\address{
  ${}^*$\
Istituto per le Applicazioni del Calcolo ``M. Picone,'' CNR I-00161 Rome, Italy
}
\address{
  ${}^\S$\
  ICRA, University of Rome ``La Sapienza,'' I-00185 Rome, Italy
  
}
\address{
${}^\P$
  INFN - Sezione di Firenze, Polo Scientifico, Via Sansone 1, 
  I-50019, Sesto Fiorentino (FI), Italy 
}
\address{
$^\dagger$
Department of Physics, University of Ferrara and INFN Sezione di Ferrara, I-44100 Ferrara,
Italy 
}

\address{
  ${}^\ddag$\
  Physics Department, University of Rome ``La Sapienza,'' I-00185 Rome, Italy
}

\address{
  ${}^{**}$\
  INFN - National Laboratories of Legnaro, I-35020 Legnaro (PD), Italy
}

\begin{abstract}
The motion of an extended body up to the quadrupolar structure is studied in the Schwarzschild background following Dixon's model and within certain restrictions (constant frame components for the spin and the quadrupole tensor, center of mass moving along  a circular orbit, etc.). We find a number of interesting situations in which deviations from the geodesic motion, due to the internal structure of the particle, can originate measurable effects.
However, the standard clock-effect for a pair co/counter-rotating bodies spinning up/down is not modified by the quadrupolar structure of the particle. 
\end{abstract}

\pacno{04.20.Cv}

\section{Introduction}

The equations of motion for an extended body in a given gravitational background were deduced
by Dixon \cite{dixon64,dixon69,dixon70,dixon73,dixon74} (hereafter the \lq\lq relativistic extended body model,"  or simply \lq\lq Dixon's model")  in multipole approximation to any order. 
In the quadrupole approximation they read
\begin{eqnarray}\fl\qquad
\label{papcoreqs1}
\frac{DP^{\mu}}{\rmd \tau_U}&=&-\frac12R^{\mu}{}_{\nu\alpha\beta}U^{\nu}S^{\alpha\beta}-\frac16J^{\alpha\beta\gamma\delta}R_{\alpha\beta\gamma\delta}{}^{;\,\mu}
\equiv F^{\rm (spin)}{}^{\mu}+F^{\rm (quad)}{}^{\mu}\ , \\
\label{papcoreqs2}
\fl\qquad
\frac{DS^{\mu\nu}}{\rmd \tau_U}&=&2P^{[\mu}U^{\nu]}+\frac43J^{\alpha\beta\gamma[\mu}R^{\nu]}{}_{\gamma\alpha\beta}\ ,
\end{eqnarray}
where $P^{\mu}=m U_p^\mu$ (with $U_p\cdot U_p=-1$) is the total four-momentum of the particle, and $S^{\mu\nu}$ is a (antisymmetric) spin tensor; 
$U$ is the timelike unit tangent vector of the \lq\lq center of mass line'' ${\mathcal C}_U$ used to make the multipole reduction, parametrized by the proper time $\tau_U$.
The tensor $J^{\alpha\beta\gamma\delta}$ is the quadrupole moment of the stress-energy tensor of the body, and has the same algebraic symmetries as the Riemann tensor. Using standard spacetime splitting techniques it can be reduced to the following form
\beq
\fl\qquad
J^{\alpha\beta\gamma\delta}=\Pi^{\alpha\beta\gamma\delta}-\bar u^{[\alpha}\pi^{\beta]\gamma\delta}-
\bar u^{[\gamma}\pi^{\delta]\alpha\beta}-3\bar u^{[\alpha}Q^{\beta][\gamma}\bar u^{\delta]}\ ,
\eeq
where $Q^{\alpha\beta}=Q^{(\alpha\beta)}$ represents the quadrupole moment of the mass distribution as measured by an observer with $4$-velocity $\bar u$. Similarly $\pi^{\alpha\beta\gamma}=\pi^{\alpha [\beta\gamma]}$ (with the additional property $\pi^{[\alpha\beta\gamma]}=0$) and 
$\Pi^{\alpha\beta\gamma\delta}=\Pi^{[\alpha\beta][\gamma\delta]}$ are essentially the body's momentum and stress quadrupole.
Moreover the various fields $Q^{\alpha\beta}$, $\pi^{\alpha\beta\gamma}$ and $\Pi^{\alpha\beta\gamma\delta}$ are all spatial (i.e. give zero after any contraction by $\bar u$). The number of independent components of $J^{\alpha\beta\gamma\delta}$ is 20: 6 in $Q^{\alpha\beta}$, 6 in $\Pi^{\alpha\beta\gamma\delta}$ and 8 in $\pi^{\alpha\beta\gamma}$. When the observer $\bar u=U_p$, i.e. in the frame associated with the momentum of the particle, 
the tensors $Q^{\alpha\beta}$, $\pi^{\alpha\beta\gamma}$ and $\Pi^{\alpha\beta\gamma\delta}$ have an intrinsic meaning.
 
There are no evolution equations for the quadrupole as well as higher multipoles as a consequence of the Dixon's construction, so their evolution is completely free, depending only on the considered body.
Therefore the system of equations is not self-consistent, and one must assume that all unspecified quantities are  known as intrinsic properties of the matter under consideration.

In order the model to be mathematically correct the following additional condition should be imposed to the spin tensor:
\beq
S^{\mu\nu}U_p{}_\nu=0.
\eeq
Such supplementary conditions (or Tulczyjew-Dixon conditions \cite{tulc59,dixon64}) are necessary to ensure the correct definition of the various multipolar terms.

Dixon's model for structured particles originated to complete and give a rigorous mathematical support to the previously introduced  Mathisson-Papapetrou model \cite{math37,papa51,cori51,pir56}, i.e. a multipole approximation to any order which includes evolution equations along the \lq\lq center line" for all the various structural quantities. The models are then different and a comparison between the two is possible at the dipolar order but not once the involved order is the quadrupole.

In this paper we limit our considerations to Dixon's model under the further simplifying assumption\cite{taub64,ehlers77} 
that the only contribution to the complete quadrupole moment $J^{\alpha\beta\gamma\delta}$ stems from the mass quadrupole moment $Q^{\alpha\beta}$, so that $\pi^{\alpha\beta\gamma}=0=\Pi^{\alpha\beta\gamma\delta}$ and
\beq
J^{\alpha\beta\gamma\delta}=-3U_p^{[\alpha}Q^{\beta][\gamma}U_p^{\delta]}\ ,\qquad Q^{\alpha\beta}U_p{}_\beta=0\ ;
\eeq

Let us introduce the  spin vector by spatial (with respect to $U_p$) duality
\beq
\label{spinvec}
S^\beta={\textstyle\frac12} \eta_\alpha{}^{\beta\gamma\delta}U_p^\alpha S_{\gamma\delta}\ ,
\eeq
where $\eta_{\alpha\beta\gamma\delta}=\sqrt{-g} \epsilon_{\alpha\beta\gamma\delta}$ is the unit volume 4-form and $\epsilon_{\alpha\beta\gamma\delta}$ ($\epsilon_{0123}=1$) is the Levi-Civita alternating symbol, 
 as well as the scalar invariant
\beq
\label{sinv}
s^2=\frac12 S_{\mu\nu}S^{\mu\nu}\ . 
\eeq
In general $s$ is not constant along the trajectory of a spinning particle. 

The assumption that the particle under consideration is a test particle means that its mass, its spin as well as its quadrupole moments must all be small enough not to contribute significantly to the background metric. Otherwise, backreaction must be taken into account.

\section{Motion of extended bodies in Schwarzschild spacetime}

Let us consider the case of the Schwarzschild spacetime, with the metric written in standard Schwarzschild coordinates,
\beq\fl\quad 
\label{metric}
\rmd  s^2 = -\left(1-\frac{2M}r\right)\rmd t^2 + \left(1-\frac{2M}r\right)^{-1} \rmd r^2 +r^2 (\rmd \theta^2 +\sin^2 \theta \rmd \phi^2),
\eeq
and introduce an orthonormal frame adapted to the static observers
\beq\fl\quad 
\label{frame}
e_{\hat t}=(1-2M/r)^{-1/2}\partial_t, \,
e_{\hat r}=(1-2M/r)^{1/2}\partial_r, \,
e_{\hat \theta}=\frac{1}{r}\partial_\theta, \,
e_{\hat \phi}=\frac{1}{r\sin \theta}\partial_\phi ,
\eeq
with dual 
\begin{equation}\fl\quad 
\omega^{{\hat t}}=(1-2M/r)^{1/2}\rmd t, \, 
\omega^{{\hat r}}=(1-2M/r)^{-1/2}\rmd r, \, 
\omega^{{\hat \theta}}=r \rmd \theta, \,
\omega^{{\hat \phi}}=r\sin \theta \rmd\phi\ .
\end{equation}
Let us assume that $U$ is  tangent to a (timelike) spatially circular orbit, hereafter denoted as the $U$-orbit, with
\beq
\label{orbita}
U=\Gamma [\partial_t +\zeta \partial_\phi ]=\gamma [e_{\hat t} +\nu e_{\hat \phi}], \qquad \gamma=(1-\nu^2)^{-1/2}\ ,
\eeq
where $\zeta$  is
the angular velocity with respect to infinity and $\Gamma$ is a normalization factor
\beq
\Gamma =\left( -g_{tt}-\zeta^2g_{\phi\phi} \right)^{-1/2}
\eeq
which assures that $U\cdot U=-1$; here dot means scalar product with respect to the metric (\ref{metric}). The angular velocity $\zeta$
is related to the local proper linear velocity $\nu$ measured in the frame (\ref{frame}) by
\beq
\zeta=\sqrt{-\frac{g_{tt}}{g_{\phi\phi}}} \nu .
\eeq
Here $\zeta$ and therefore also $\nu$ are assumed to be constant along the $U$-orbit.  
We limit our analysis to  the equatorial plane ($\theta=\pi/2$) of the Schwarzschild solution; as a convention, the physical (orthonormal) component along $-\partial_\theta$, perpendicular to the equatorial plane will be referred to as along the positive $z$-axis and will be indicated by $\hat z$, when necessary.

Among the circular orbits particular attention is devoted to the co-rotating $(\zeta_+)$
 and counter-rotating $(\zeta_-)$ timelike circular geodesics, having respectively 
$\zeta_\pm\equiv \pm\zeta_K=\pm (M/r^3)^{1/2}$, so that 
\beq\fl\quad 
\label{Ugeos}
U_\pm=\gamma_K [e_{\hat t} \pm \nu_K e_{\hat \phi}]\ , \qquad \nu_K=\left[\frac{M}{r-2M}\right]^{1/2}, \qquad \gamma_K=\left[\frac{r-2M}{r-3M}\right]^{1/2}\ ,
\eeq
with the timelike condition $\nu_K < 1$ satisfied if $r>3M$.
It is convenient to introduce also the Lie relative curvature \cite{idcf1,idcf2} of each orbit 
\beq
k_{\rm (lie)}=-\partial_{\hat r} \ln \sqrt{g_{\phi\phi}}=-\frac1r\left(1-\frac{2M}{r}\right)^{1/2}=-\frac{\zeta_K}{\nu_K}\ .
\eeq
 
Let $P=m\, U_p$ such that
\begin{equation}
\label{Ptot}
U_p=\gamma_p\, [e_{\hat t}+\nu_p e_{\hat \phi}]\ , \quad \gamma_p=(1-\nu_p^2)^{-1/2}\ , 
\end{equation}
i.e. let us assume that $U_p$ also is tangent to a circular orbit and
set up an orthonormal frame adapted to $U_p$ given by
\beq
\fl\qquad
e_0=U_p\ , \qquad e_1=e_{\hat r}\ , \qquad e_2=\gamma_p\, [\nu_p e_{\hat t}+ e_{\hat \phi}]\ , \qquad e_3=e_{\hat z}\ ;
\eeq
hereafter all frame components of the various fields  are  meant to be referred to such a frame.
Note that the assumption of having both $U$ and $U_p$ aligned with a circular orbit certainly represents a restriction to our analysis, but leads to great simplifications to Dixon's equations.
On the other hand, dealing with the model in its complete generality, i.e. with $U_p$ having no relation a priori with $U$, is a hard task to pursue analytically due to its mathematical complexity, so that a numerical analysis would be needed.

The spin vector is orthogonal to $U_p$, so that 
\beq
S=S^1e_1+S^2e_2+S^3e_3\ .
\eeq
Furthermore we have 
\beq
Q_{00}=Q_{01}=Q_{02}=Q_{03}=0\ .
\eeq
We also assume that $S^1=0=S^2$, and that the remaining component $S^3=S^{\hat z}=s$ (mimic of a particle moving on a circular orbit and spinning around the $z$-axis) as well as the surviving components of the mass quadrupole moment are all constant along the path.  The latter  assumption corresponds to the definition of \lq\lq quasirigid motion" (or \lq\lq quasirigid bodies") due to Ehlers and Rudolph \cite{ehlers77}.  
Clearly in a more realistic situation the latter hypothesis should be released.

Consider first the evolution equations (\ref{papcoreqs2}) for the spin tensor. They imply that 
\beq
Q_{12}=Q_{13}=Q_{23}=0\ ,
\eeq
and introducing the following \lq\lq structure functions" of the extended body
\beq
Q_{11}=Q_{33}+f, \qquad Q_{22}=Q_{33}+f', 
\eeq
they also give
\beq
\label{moto1}
0=(\nu\nu_p-\nu_K^2)s+m\frac{\nu_K}{\zeta_K}(\nu-\nu_p)+3\nu_K\zeta_K\frac{\gamma_p\nu_p}{\gamma}\, f\ .
\eeq
Consider then the equations of motion (\ref{papcoreqs1}). 
They imply that
\beq\fl\quad
\label{moto2old}
0=\nu_K\zeta_K(2\nu_p+\nu)s-\frac32\frac{\zeta_K^2}{\gamma \gamma_p}[-f'+\gamma_p^2(2+\nu_p^2)f]+m(\nu\nu_p-\nu_K^2)\ .
\eeq
After eliminating $s$ through Eq. (\ref{moto1}), the previous equation becomes 
\begin{eqnarray}
\label{moto2}\fl\quad
0&=&(\nu\nu_p-\nu_K^2)(-f'+2\gamma_p^2f)
+\gamma_p^2[2\nu\nu_p \nu_K^2 +
\nu_p^2(\nu\nu_p+\nu_K^2)+2\nu_K^2\nu_p^2]f\nonumber\\
\fl
&&-\frac23\frac{m\gamma \gamma_p}{\zeta_K^2}
(\nu^2\nu_p^2+2\nu_p^2\nu_K^2-3\nu\nu_p\nu_K^2-\nu^2\nu_K^2+\nu_K^4)\ ,
\end{eqnarray}
where 2$^{\rm nd}$ and 4$^{\rm nd}$ order polynomial expressions in the velocities have been collected. 
Solving the last two equations for $\nu$ and $\nu_p$ in terms of $s$ and $f$, $f'$ completely determines the motion.

The  quadrupole moment tensor of a mass distribution of density $\mu$ is defined classically  by
\beq
\label{Qdef}
Q_{\rm (class)}^{ab}=\int_V \mu \, (3x^ax^b-r^2\delta^{ab})d^3x\qquad a=1,2,3
\eeq
with $r^2=\delta_{ab}x^ax^b$ and it is tracefree. It is therefore natural to assume the same property holding for the relativistic quadrupole moment tensor, obtaining
\beq
\fl\qquad
0=Q_{11}+Q_{22}+Q_{33}=Q_{33}+f+Q_{33}+f'+Q_{33}=3Q_{33}+f+f'\ ,
\eeq
so that the components  $Q_{ab}$ in this case are completely determined by the two \lq\lq structure functions"
$f$ and $f'$
\begin{eqnarray}
\fl\qquad Q_{11}&=&\frac23 f-\frac13 f'\ , \quad Q_{22}=-\frac13f+\frac23 f'\ , \quad
Q_{33}=-\frac13 (f+f')\ .
\end{eqnarray}
If the body is axially symmetric about the $z$-axis, then $f=f'$ and the frame components of $Q$ reduce to
\beq
Q_{ab}={\rm diag}\,[f/3,f/3,-2f/3]\ .
\eeq
Without entering the problems of a relativistic definition of the quadrupole moment tensor $Q_{ab}$ generalizing Eq. (\ref{Qdef}) in terms of integrals over the volume of the body itself (the correct procedure is outlined in Dixon's works \cite{dixon64,dixon69,dixon70,dixon73,dixon74}), 
we note that $Q_{ab}$ (only defined all along the world line with tangent vector $U$) can be interpreted as the 
the mass quadrupole moment of the extended body as measured by the observer $U_p$.

It is quite natural to introduce the following rescaled dimensionless angular and quadrupolar momentum quantities
\beq
\label{adim}
\sigma=\frac{s}{m}\zeta_K\ , \quad F=\frac{f}{m}\zeta_K^2\ , \quad F'=\frac{f'}{m}\zeta_K^2\ ,
\eeq
due to the fact that along a circular orbit $r=\,$const. 
The quantities $\sigma$, $F$ and $F'$ are necessarily small. Although the quadrupolar terms $f$ and $f'$ are small only for a quasi-spherical body, the further rescaling by $\zeta_K=(M/r^3)^{1/2}$ makes indeed them small in any case. In fact, the radius of the orbit is assumed to be large enough in comparison with certain natural length scales like $|s|/m$ (also known as the M\o ller radius \cite{mol} of the body), $(|f|/m)^{1/2}$, $(|f'|/m)^{1/2}$ associated with the body itself in order to avoid backreaction effects.  
Furthermore, we also require that the characteristic length associated with the quadrupole are small if compared to the M\o ller radius of the body, in order that the multipolar expansion of the body's stress-energy tensor underlying the Dixon's model be consistent.

Eqs. (\ref{moto1}) and (\ref{moto2}) then become
\beq
\label{moto1_bis}
0=(\nu\nu_p-\nu_K^2)\sigma+\nu_K(\nu-\nu_p)+3\nu_K\frac{\gamma_p\nu_p}{\gamma}\, F\ ,
\eeq
and
\begin{eqnarray}\fl\quad
\label{moto2_bis}
0&=&(\nu\nu_p-\nu_K^2)(-F'+2\gamma_p^2F)
+\gamma_p^2[2\nu\nu_p \nu_K^2 +
\nu_p^2(\nu\nu_p+\nu_K^2)+2\nu_K^2\nu_p^2]F\nonumber\\
\fl
&&-\frac23\gamma \gamma_p
(\nu^2\nu_p^2+2\nu_p^2\nu_K^2-3\nu\nu_p\nu_K^2-\nu^2\nu_K^2+\nu_K^4)\ .
\end{eqnarray}
The above relations define the kinematical conditions allowing circular motion of the extended body taking into account its spinning and quadrupolar structures. 
The contribution due to the spinning structure disappears (i.e. remains arbitrary) when $\nu\nu_p=\nu_K^2$, i.e. if $\nu_K$ is taken to be the geometrical mean of $\nu$ and $\nu_p$. In fact, in this case, we have
\beq
\label{moto1_tris}
F=-\frac{\gamma(\nu-\nu_p)}{3  \gamma_p\nu_p}\ ,
\eeq
and
\beq
\label{moto2_tris}
0=(\nu_K^2 +2\nu_p^2)(\nu-\nu_p)+
(2\nu_p^2-\nu^2-\nu_K^2)\nu_p\ .
\eeq
After some manipulation using the condition $\nu\nu_p=\nu_K^2$ the latter equation turns out to be identically satisfied, whereas Eq. (\ref{moto1_tris}) becomes 
\beq
\label{moto1_tris2}
F=-\frac{\gamma}{3 \nu_K^2}\sqrt{1-\frac{\nu_K^4}{\nu^2}}(\nu^2-\nu_K^2)\ ,
\eeq
with $F'$ arbitrary.
 
Equations (\ref{moto1_bis}) and (\ref{moto2_bis}) can be examined in other special cases; for example:
\begin{enumerate}
\item \underline{$\sigma=0,\, F\not =0, F'\not =0$}.
\begin{eqnarray}
\fl
0&=&\nu-\nu_p+3\frac{\gamma_p\nu_p}{\gamma}\, F\ ,\nonumber\\
\fl
0&=&(\nu\nu_p-\nu_K^2)(-F'+2\gamma_p^2F)
+\gamma_p^2[2\nu\nu_p \nu_K^2 +
\nu_p^2(\nu\nu_p+\nu_K^2)+2\nu_K^2\nu_p^2]F\nonumber\\
\fl
&&-\frac23\gamma \gamma_p
(\nu^2\nu_p^2+2\nu_p^2\nu_K^2-3\nu\nu_p\nu_K^2-\nu^2\nu_K^2+\nu_K^4)\ .
\end{eqnarray}
We notice that if $F=0$ these equations imply $\nu=\nu_p$ and
\beq
0=(\nu_p^2-\nu_K^2)\left[F'
+\frac23\gamma_p^2(\nu_p^2-\nu_K^2)\right]\ ,
\eeq
that is $\nu_p=\pm \nu_K,$
or if $\nu_p^2\not=\nu_K^2$,
\beq
F'=-\frac23\gamma_p^2(\nu_p^2-\nu_K^2)\ ,
\eeq
allowing nongeodesic motion only due to the quadrupole moment tensor.
\item \underline{$\sigma\not =0,\, F=0, F'\not =0$}.
\begin{eqnarray}
\fl
0&=&(\nu\nu_p-\nu_K^2)\sigma+\nu_K(\nu-\nu_p),\nonumber\\
\fl
0&=&(\nu\nu_p-\nu_K^2)F'
+\frac23\gamma \gamma_p
(\nu^2\nu_p^2+2\nu_p^2\nu_K^2-3\nu\nu_p\nu_K^2-\nu^2\nu_K^2+\nu_K^4)\ .
\end{eqnarray}

\item \underline{$\sigma\not=0,\, F\not =0, F'=0$}.
\begin{eqnarray}
\fl
0&=&(\nu\nu_p-\nu_K^2)\sigma+\nu_K(\nu-\nu_p)+3\nu_K\frac{\gamma_p\nu_p}{\gamma}\, F\ ,\nonumber\\
\fl
0&=&\gamma_p^2[2\nu\nu_p(1+\nu_K^2) + \nu_p^2(\nu\nu_p+3\nu_K^2)-2\nu_K^2]F
\nonumber\\
\fl&&
-\frac23\gamma \gamma_p
(\nu^2\nu_p^2+2\nu_p^2\nu_K^2-3\nu\nu_p\nu_K^2-\nu^2\nu_K^2+\nu_K^4)\ .
\end{eqnarray}
\end{enumerate}
The case of a spinning particle with vanishing quadrupole moment tensor, i.e. $F=0=F'$ has been already studied in \cite{bdfgschwclock}.

Let us turn to Eqs. (\ref{moto1_bis})--(\ref{moto2_bis}) and investigate the case of extended bodies with internal structure (dipolar and quadrupolar) compatible with a nearly geodesic motion. 
In the case of vanishing quadrupole moments Eqs. (\ref{moto1_bis})--(\ref{moto2_bis}) reduce to  \cite{bdfgschwclock} 
\begin{eqnarray}
0&=&(\nu\nu_p-\nu_K^2)\sigma+\nu_K(\nu-\nu_p)\ ,  \nonumber\\
0&=&\nu^2\nu_p^2+2\nu_p^2\nu_K^2-3\nu\nu_p\nu_K^2-\nu^2\nu_K^2+\nu_K^4\ .
\end{eqnarray}
In the limit of small spin $\sigma$ we find
\begin{equation}
\label{nu_spin}
\nu=\pm\nu_K-\frac32\nu_K \sigma+O(\sigma^2)\ , \qquad
\nu_p=\nu+O(\sigma^2)\ .
\end{equation}

If the contribution of quadrupolar terms can be considered negligible with respect to the dipolar ones and comparable with second order terms in the spin itself, one can consider corrections to Eq. (\ref{nu_spin}) as given by
\beq
\fl\quad
\nu_\pm\simeq\pm\nu_K-\frac32 \nu_K \sigma \pm\frac38 (2F+\sigma^2)\nu_K\ , \qquad \nu_p^{(\pm)}\simeq\nu_\pm \pm 3 (F-\sigma^2)\nu_K\ ,
\eeq
where the signs $\pm$ correspond to co/counter rotating orbits.
The corresponding angular velocity $\zeta_\pm=(\zeta_K/\nu_K)\nu_\pm$ and its reciprocal are
\begin{eqnarray} 
\label{zetasol}
\zeta_\pm&\simeq& \pm\zeta_K \left[1\mp\frac32 \sigma + \frac38 (2F+\sigma^2)\right]\ , \nonumber\\ \frac1{\zeta_\pm}&\simeq& \pm\frac{1}{\zeta_K}+\frac{3}{2\zeta_K} \sigma \pm \frac{3}{8\zeta_K}(5\sigma^2-2F)\ .
\end{eqnarray}
Furthermore, the period of revolution around the central source will consist of three different terms
\beq
\label{periodo}
T=\frac{2\pi}{|\zeta_\pm|}=T_K\,|1\pm\lambda_d+\lambda_q\,|\ , 
\eeq
where 
\beq
\label{per1}
T_K=\frac{2\pi}{\zeta_K}\ , \qquad
\lambda_d=\frac{3}{2} \sigma\ , \qquad
\lambda_q=\frac{3}{8}(5\sigma^2-2F)\ .
\eeq
A direct measurement of $T$ will then allow to estimate the quantity $F$ determining the quadrupolar structure of the body, if its spin is known.
Note that the fraction $\lambda_d$ due to the spin is different depending on whether the body is spinning up or down, whereas the term $\lambda_q$ due to the quadrupole has a definite sign once the shape of the body is known ($F$ cannot change its sign). 

In the case of the Earth the nondimensional quantities (\ref{adim}) turn out to be given by $\sigma\approx2.3\times10^{-15}$ and $F=F'\approx-1.8\times10^{-20}$, since $s/m_\oplus\approx3.4\times10^2$ cm and $f=f'=-J_2m_\oplus r_\oplus^2$, with $J_2\approx10^{-3}$, and the distance between the Earth and the Sun is $r\approx1.5\times10^{13}$ cm.
The correction to the geodesic value $T_K\approx9.425\times10^{17}$ cm due to the spin is $\lambda_d\approx3.4\times10^{-15}$, whereas the correction due to the quadrupole turns out to be $\lambda_q\approx1.3\times10^{-20}$ as from Eq. (\ref{per1}).

An interesting opportunity to test the quadrupole effect of extended body would arise from the motion of a binary pulsar system around Sgr A$^*$, the supermassive ($M\simeq 10^6\ M_\odot$) black hole located at the Galactic Center \cite{falcke,muno}. To illustrate the order of magnitude of the effect, we may consider the binary pulsar system PSR J0737-3039 as orbiting Sgr A$^*$ at a distance of $r\simeq 10^9$ Km. 
The PSR J0737-3039 system consists of two close neutron stars (their separation is only $d_{AB} \sim 8 \times 10^5$ Km) of comparable masses $m_A\simeq 1.4\ M_\odot$, $m_B \simeq 1.2\ M_\odot$), but very different intrinsic spin period ($23$ ms of pulsar A vs $2.8$ s of pulsar B) \cite{lyne}. 
Its orbital period is about $2.4$ hours, the smallest yet known for such an object. Since the intrinsic rotations are negligible with respect to the orbital period, we can treat the binary system as a single object with reduced mass $\mu_{AB} \simeq 0.7 \ M_\odot$ and intrinsic rotation equal to the orbital period. 
The spin parameter thus turns out to be equal to $\sigma\approx6\times10^{-8}$, whereas the quadrupolar parameters are $F=F'\approx9.6\times10^{-10}$, since we have taken $f=f'=\mu_{AB}d_{AB}^2$ as a rough estimate.
The correction to the geodesic value $T_K\approx1.6\times10^{16}$ cm due to the spin is $\lambda_d\approx9\times10^{-8}$, whereas the correction due to the quadrupole turns out to be $\lambda_q\approx-7.2\times10^{-10}$ as from Eq. (\ref{per1}).

The black hole at the Galactic Center is actually expected to be rotating \cite{genzel}. Therefore, a more detailed analysis would take into account the effect of rotation of Sgr A$^*$ on the estimate of the various contributions to the period of revolution of the PSR J0737-3039 binary system. 
We will discuss such an extension of the present treatment in a forthcoming paper.

\section{Quadrupolar corrections to the clock-effect for spinning test particles}

Bini, de Felice and Geralico have investigated in \cite{bdfgschwclock} the gravitomagnetic clock effect appearing for oppositely orbiting both spin-up or spin-down  particles in the Schwarzschild spacetime.
They found that spinning test particles move on circular orbits which, to first order in the spin parameter $\sigma$, 
are close to a geodesic, with
\beq
\label{clock}
\frac{1}{\zeta_{(\pm,\pm)}}=\pm \frac{1}{\zeta_K} \pm \frac{3}{2\zeta_K} |\sigma|\ ,
\eeq 
where the signed magnitude $\sigma=\pm|\sigma|$ of the spin parameter has been introduced.
The signs in front of $1/\zeta_K$ correspond to co/counter-rotating 
orbits while the signs in front of $|\sigma|$ refer to a positive or negative spin direction along the $z$-axis; for instance, the quantity 
$\zeta_{(+,-)}$ denotes the angular velocity of $U$ corresponding to a co-rotating orbit $(+)$ with spin-down $(-)$ alignment, etc.
Therefore one can measure the difference in the arrival times after one complete revolution with respect to a 
static observer. The coordinate time difference in given by: 
\beq\label{deltat}
\Delta t_{(+,+;-,+)}= 2\pi \left(\frac{1}{\zeta_{(+,+)}}+\frac{1}{\zeta_{(-,+)}}\right)= \frac{6 \pi}{\zeta_K}|\sigma|\ ,
\eeq
and analogously for $\Delta t_{(+,-;-,-)}$. 

Let us analyze now the introduction of quadrupolar terms.
Eq. (\ref{zetasol}) implies that Eq. (\ref{clock}) becomes
\beq
\label{clock2}
\frac{1}{\zeta_{(\pm,\pm)}}=\pm \frac{1}{\zeta_K} \pm \frac{3}{2\zeta_K}|\sigma|\pm \frac{3}{8\zeta_K}(5\sigma^2-2F)\ ,
\eeq 
so that the last term does not contribute to the clock effect, since the $\pm$ signs in front of it correspond to co/counter-rotating orbits (like those in front of $1/\zeta_K$), and thus cancels anyway.
Therefore, no modifications are induced by quadrupolar terms in Eq. (\ref{deltat}). 

\section{Concluding remarks}

We have studied the motion of quadrupolar particles on a Schwarzschild background following Dixon's model. 
In the simplified situation of constant frame components (with respect to a natural orthonormal frame) of both the spin and the quadrupole tensor of the particle we have found  the kinematical conditions to be imposed to the particle's structure in order the orbit of the particle itself be  circular and confined on the equatorial plane.
Co-rotating and counter-rotating particles result to have a non-symmetric speed in spite of the spherical symmetry of the background,  due to their internal structure.
This fact has been anticipated when studying spinning particles only, i.e. with vanishing quadrupole moments \cite{bdfgschwclock}.

We have then discussed the modifications due to the quadrupole which could be eventually observed in experiments. Such experiments, however, cannot  concern standard clock effects, because we have shown that there are no contributions arising from the quadrupolar structure of the body in this case. 
In contrast, the effect of the quadrupole terms could be important when considering the period of revolution of an extended body around the central source. Measuring the period will provide an estimate of the quantities $F$, $F'$ determining the quadrupolar structure of the body, if its spin is known. On the other hand, the complete knowledge of the internal structure of the body will allow to estimate the period of revolution. In the latter case, the comparison between the measured period (known from observations) with the value predicted by the Dixon's model could provide a test for the model itself.  

It would be of great interest to extend this analysis to systems with varying quadrupolar structure and emitting gravitational waves without perturbing significantly the background spacetime. 
Usually variable quadrupole moment is generated in a test astronomical body of mass $m$ because of tides produced by the central source of mass $M\gg m$. Now the net gravitational radiation due to motion of $m$ is due to its orbit around $M$, the time varying tides and the interference between 
these two \cite{mash1,mash2}. 
We deserve such an investigation to future works.

\end{document}